\documentclass[lettersize,journal]{IEEEtran}
\usepackage{mathtools}
\usepackage[utf8x]{inputenc}

\usepackage{makecell}
\usepackage{color,soul}
\definecolor{purple}{rgb}{0.5,0,0.5}
\usepackage{subfigure}
\usepackage{moreverb}
\usepackage{epsfig}
\usepackage{amsmath,amssymb,bm,amsthm,mathrsfs,dsfont}
\usepackage{fancyhdr}
\usepackage{adjustbox,lipsum}
\usepackage[font={small}]{caption}
\usepackage{xcolor,soul}
\usepackage{graphicx}
\usepackage{amsfonts}
\usepackage{cite}
\usepackage{microtype}
\usepackage[colorlinks,bookmarksopen,bookmarksnumbered,citecolor=blue,urlcolor=red]{hyperref}

\usepackage{pgfplots}
\pgfplotsset{compat=1.18}
\usepackage{xcolor}
\definecolor{darkGreen}{rgb}{0.47,0.67,0.19}
\definecolor{saddleBrown}{rgb}{0.55,0.27,0.07}
\definecolor{pinkDark}{rgb}{0.72,0.27,1.00}

\usepackage{algorithmicx}
\usepackage{algpseudocode}
\usepackage[norelsize, linesnumbered, ruled, vlined, lined, boxed, commentsnumbered]{algorithm2e}
\usepackage{booktabs}

\usepackage{amsmath,amssymb}
\usepackage{tikz}
\usepackage[mode=buildnew]{standalone}
\allowdisplaybreaks
\usepackage[nodisplayskipstretch]{setspace} 
\begin{document}
\title{ %\huge
%\\
Temporal vs Spectral Resolution in CSI Reporting for Wi-Fi ISAC 
%\\ \textcolor{red}{Author lists}
%\red{workign title ....}
}
\author{\IEEEauthorblockN{
 %1\textsuperscript{st} 
 Abolfazl Zakeri\IEEEauthorrefmark{1},~Sumit Roy\IEEEauthorrefmark{2}, Nhan~Thanh~Nguyen\IEEEauthorrefmark{1}, and Markku Juntti\IEEEauthorrefmark{1}}
 %\vspace{-1 em }
 \\
 \IEEEauthorrefmark{1}\normalsize 
 Centre for Wireless Communications,  email:\{abolfazl.zakeri,\,nhan.nguyen,\,markku.juntti\}@oulu.fi
\\\  \IEEEauthorrefmark{2}\normalsize 
University of Washington, Seattle, USA, email: sroy@uw.edu
\vspace{-2.5 em}
}

\maketitle
\begin{abstract}
 CSI feedback is fundamental to proposed uses of Wi-Fi sensing such as human activity recognition (HAR); but it also constitutes significant overhead as resources must be allocated away from data communications. Wi-Fi CSI extraction can be explored along two dimensions in IEEE 802.11bf: reporting intensity, i.e., \textit{temporal resolution}, and number of reported subcarriers, i.e., \textit{spectral resolution}. We explore periodic and IEEE 802.11bf threshold-based reporting at matched rates, and uniform decimation and block-mean grouping at matched report sizes. We further propose reporting the mean CSI accumulated since the last report.
Using the 802.11bf report size as the metric, we evaluate the accuracy--cost tradeoffs on three public datasets with our proposed model, a
convolutional neural network combined with a gated recurrent unit and temporal
pyramid pooling.
The proposed periodic reporting with mean payload and block-mean grouping improves accuracy by up to $53$ percentage points at equal CSI overhead cost. Temporal resolution is generally more cost-effective than spectral resolution. With only at most $7.7\%$ of full-CSI feedback cost, the proposed scheme achieves $96$--$98\%$ of full-CSI accuracy.
\end{abstract}

\vspace{-1 em}
\section{Introduction}\label{sec_intor}
A  significant milestone for commercial deployment of Wi-Fi-sensing was the adoption of IEEE 802.11bf-2025 amendment. This enables device-free sensing by defining sensing procedures initiated by network nodes to measure channel state information (CSI) and its corresponding feedback by the responder. Use cases of this integrated sensing and communication (ISAC) feature include motion and presence detection, human activity recognition
(HAR), elder-care monitoring, and gesture recognition
~\cite{wba2019sensing,du2025overview}. A prominent home ISP vendor such as Comcast launched camera-free Wi-Fi motion sensing through its Xfinity gateways
in August 2026~\cite{comcast2026shield}, and Wi-Fi~8 devices are expected to
support sensing based on IEEE 802.11bf~\cite{IEEE80211bf2025,intel2025wifi8}.
\\\indent 
Many of these applications are based on detailed CSI, which captures how the wireless
channel varies across subcarriers and over time. In a trigger-based (TB) version of sensing and CSI acquisition in IEEE 802.11bf, the access point (AP) initiates channel sounding and the
station (STA) returns a sensing measurement report~\cite{IEEE80211bf2025}.
Depending on the antenna configuration, channel bandwidth, and subcarrier
grouping, a CSI report can range from 44 to 64\,624 octets~\cite[Sec.~9.4.1.81]{IEEE80211bf2025}. HAR requires frequent temporal CSI
measurements by the client that must be fed back, constituting a major source of sensing overhead. Such channel
sounding and CSI reporting consume airtime that would otherwise be available for
data transmission, thereby reducing communication throughput~\cite{sahoo2024sensing}. Reducing CSI feedback is thus central to effectively
integrating sensing into  Wi-Fi networks that are primarily built for communication~\cite{du2025overview}.
\\\indent 
IEEE 802.11bf provides two mechanisms for controlling CSI feedback overhead: threshold-based reporting (impacts temporal resolution) and subcarrier grouping (impacts spectral resolution). Threshold-based reporting triggers a report only when CSI variation exceeds a configured threshold~\cite[Sec.~11.55.1.5.2.6.2]{IEEE80211bf2025}, while subcarrier grouping reduces report size by retaining a subset of subcarriers according to a configured grouping factor~\cite[Sec.~9.4.1.81.3]{IEEE80211bf2025}.
\\\indent 
% IEEE 802.11bf provides two mechanisms that directly control CSI feedback: (1) threshold-based reporting and (2) subcarrier grouping, which control the temporal and spectral resolution of the CSI available to the sensing application. 
% Threshold-based reporting allows an STA to transmit a report only when the CSI
% variation exceeds a configured threshold, thereby reducing the reporting
% frequency~\cite[Sec.~11.55.1.5.2.6.2]{IEEE80211bf2025}. Subcarrier grouping
% reduces the report size by retaining a subset of subcarriers according to a
% configured grouping factor~\cite[Sec.~9.4.1.81.3]{IEEE80211bf2025}. %These mechanisms control the temporal and spectral resolution of the CSI available to the sensing application. 
Only a handful of prior studies have examined the cost or resource implications of CSI reporting in Wi-Fi sensing~\cite{zakeri2026age, varshney2023adaptive}. On the temporal side,
Varshney \textit{et al.}~\cite{varshney2023adaptive} proposed adaptive
thresholds for CSI reporting and evaluated range and velocity estimation at
mmWave frequencies, but did not compare threshold-based
periodic reporting at matched average rates. Meneghello \textit{et
al.}~\cite{meneghello2023isac} varied the sounding bandwidth and sampling
period for HAR on commercial IEEE 802.11ax devices, but did not account for
CSI feedback cost. On the spectral side, Sahoo \textit{et al.}~\cite{sahoo2024sensing}
used the 802.11bf report size to quantify airtime and throughput loss, but did
not evaluate sensing accuracy. Subcarrier-selection methods~\cite{singh2025slimsense,behzad2025subcarrier} retain task-relevant
subcarriers, while learned compression methods~\cite{yang2022efficientfi,cerutti2025efficient} can substantially reduce CSI
size; however, they rely on index sets or decoders outside the standardized
reporting mechanisms considered~here. 
%To the best of our knowledge, no prior work jointly evaluates the standard's temporal and spectral reporting mechanisms under a common feedback-cost metric and recognition accuracy.
\\\indent 
In this paper, we study CSI feedback for Wi-Fi based HAR along two dimensions: i) reporting
intensity, i.e., temporal resolution, and ii) number of reported subcarriers, i.e., spectral resolution. We compare
periodic, random, and threshold-based reporting at matched average rates,
with each report carrying either the current CSI measurement or the mean of
the measurements accumulated since the previous report. For spectral
reduction, we consider uniform decimation, as specified by the standard, and
block-mean averaging over adjacent subcarriers. The AP reconstructs the CSI
sequence using zero-order hold and performs HAR classification with a neural
network. We evaluate each configuration in the accuracy--cost plane using
the 802.11bf report size as the feedback-cost metric and characterize the
resulting Pareto frontiers on three public datasets: SignFi~\cite{ma2018signfi}, CSI-HAR~\cite{moshiri2021csi}, and NTU-Fi~\cite{yang2023sensefi}. Our main contributions
are as follows.
\vspace{-.5 em}
\begin{itemize}
  \item We develop a unified overhead cost model for CSI reporting in HAR
  tasks that captures both temporal reporting and spectral reduction, and
  propose periodic reporting with mean payload and block-mean subcarrier
  grouping alongside 802.11bf threshold-based reporting.
  \item We propose a CNN--GRU with a temporal-pyramid readout that supports
  arbitrary numbers of reported subcarriers and temporal lengths, enabling
  one architecture across reporting configurations. With full CSI, it outperforms the models reported with each dataset (SignFi CNN~\cite{ma2018signfi}, SenseFi LeNet~\cite{yang2023sensefi}, and CSI-HAR 2D-CNN~\cite{moshiri2021csi})
by $6.5$, $2.7$, and $5.5$ percentage points, respectively.
  % With full CSI, it
  % improves accuracy over the baseline models by $6.5$, $2.7$, and $5.5$
  % percentage points on SignFi, NTU-Fi HAR, and CSI-HAR, respectively.
  \item We characterize the accuracy--cost Pareto frontiers across the three
  datasets. The results suggest that 
  %showing that periodic reporting is generally more cost-efficient
  %than threshold-based reporting and that 
  temporal resolution is more valuable than spectral resolution. With at most $7.7\,\%$ of the full-CSI
  feedback cost, the proposed scheme reaches $95.7$--$98.3\,\%$ of full-CSI
  accuracy.
\end{itemize}

\allowdisplaybreaks
\section{System Model and Problem Formulation}\label{sec:model}
%\subsection{Sensing Session}
We consider a TB sensing session in IEEE 802.11bf~\cite{IEEE80211bf2025} between a sensing initiator, the AP, and a sensing responder, the station (STA). A sensing measurement exchange session comprises \(T\) sounding \textit{instances} indexed by \({t\in\mathcal{T}=\{0,\dots,T-1\}}\).\footnote{The number of instances is determined primarily by the sensing-window duration, i.e., how long sensing/channel observation is performed, and the sensing sampling rate. The index \(t\) denotes a sounding instance rather than an exact CSI measurement timestamp.} At each instance, the AP transmits a null data packet (NDP), and the STA measures the channel state information (CSI). We denote the CSI instance $t$ by \(\mathbf{X}(t)\in\mathbb{C}^{N\times K}\), comprising \(N=N_{\mathsf{T}}N_{\mathsf{R}}\) total of Tx-Rx (spatial) links and
\(K\) subcarriers, where \(N_{\mathsf{T}}\) and \(N_{\mathsf{R}}\) are the numbers of transmit and receive antennas, respectively.
%antenna links and \(K\) subcarriers, where \(N_{\mathsf{T}}\) and \(N_{\mathsf{R}}\) denote the numbers of transmit and receive chains, respectively.
\\\indent 
Each session is associated with a label \(y\in\mathcal{Y}\), e.g., an activity or gesture, and the sensing machine-learning application performs %the task
an end task after collecting \(T\) CSI measurements/samples, which need not be uniformly spaced in time. Sending every \(\mathbf{X}(t)\) back to the AP as full CSI, however, incurs substantial feedback (reporting) overhead. The goal is therefore to reduce CSI feedback along two dimensions: \emph{when} the STA reports (temporal resolution) and \emph{how many subcarriers} each report contains (spectral resolution). Both mechanisms are supported by IEEE 802.11bf~\cite{IEEE80211bf2025}, yet have received limited investigation in the literature. The work in \cite{varshney2023adaptive} investigates temporal resolution, but only from a threshold-based~perspective.
\vspace{-1 em}
\subsection{Temporal Resolution Modeling}\label{sec:temporal}
Let $\beta(t)\in\{0,1\}$ indicate whether the STA sends a report of sounding instance
$t$, and let $t_1<\dots<t_m<\dots<t_M$ be the \textit{report} instants, with
$t_1=0$ and $M=\sum_{t\in\mathcal{T}}\beta(t)$ being the number of the reported
CSI measurements. The average reporting \textit{rate}, denoted by $\bar \alpha$, is
\begin{equation}
\begin{aligned}
    \bar{\alpha} = \dfrac{\mathbb{E}[M]}{T} \in [0,1].
    \label{eq:rate}
    \end{aligned}
\end{equation}
Let $\tilde{\mathbf{X}}(t_m)$ denote the CSI report of instance $t_m$. A report
carries either the current frame, referred to as ``instant'', or the ``mean'' of
all frames since the last report:
\begin{align}
    \tilde{\mathbf{X}}(t_m)=
    \begin{cases}
      \mathbf{X}(t_m), & \text{instant},\\[2pt]
      \dfrac{1}{L_m}\displaystyle\sum_{t=t_{m-1}+1}^{t_m}\mathbf{X}(t), & \text{mean},
    \end{cases}
    \label{eq:payload}
\end{align}
where $L_m=t_m-t_{m-1}$ and $t_0=-1$. 
The instant payload is the default in the standard; the proposed mean payload has
the same report size and hence the same feedback cost. We study three rules for $\beta(\cdot)$:
\\
$\bullet~$\emph{Periodic:} $\beta(t)=\mathds{1}\{t\bmod u=0\}$ with
  $u:=\mathsf{round}(1/\bar{\alpha})$, so that $M=\lceil T/u\rceil\approx\bar{\alpha} T$ in
  every session.
  \\
$\bullet~$\emph{Random:} $\beta(0)=1$ and $\beta(t)\sim\mathsf{Bernoulli}(p)$,
  independent over $t\geq1$, with $p=(\bar{\alpha} T-1)/(T-1)$, so that
  $\mathbb{E}[M]=\bar{\alpha} T$. It models report loss or contention with no memory,
  at the same mean rate.
  \\
$\bullet~$\emph{Threshold-based:} this follows the CSI reporting rule of
802.11bf~\cite[Sec.~11.55.1.5.2.6.2]{IEEE80211bf2025}. At every instance, the STA
sends a short report with a CSI variation value $v(t)$, and the AP requests the full
report only if $v(t)\geq\tau$, for a threshold $\tau\in[0,1]$. The standard leaves
$v(t)$ to the implementation; we use
\begin{equation}
    v(t)=1-\frac{\big|\langle \mathbf{X}(t),\mathbf{X}(\hat{t})\rangle_{\mathsf{F}}\big|}
    {\|\mathbf{X}(t)\|_{\mathsf{F}}\,\|\mathbf{X}(\hat{t})\|_{\mathsf{F}}},
    \label{eq:variation}
\end{equation}
where $\hat{t}=\max\{t_m: t_m<t\}$ is the last report instant and
$\langle\mathbf{X},\mathbf{Y}\rangle_{\mathsf{F}}=\mathsf{tr}(\mathbf{X}^{\mathsf{H}}\mathbf{Y})$
is the Frobenius inner product over the $N$ links and $K$ subcarriers. By the
Cauchy--Schwarz inequality, $v(t)\in[0,1]$; it is invariant to a common phase
rotation. % and to a gain change of $\mathbf{X}(t)$.
\\\indent 
The standard quantizes $\tau$ to $\{0,0.1,\dots,1\}$~\cite[Table~9-417z]{IEEE80211bf2025}
but leaves the mapping from channel change to $v(t)$ to the implementation; any
increasing rescaling of $v(t)$ moves our threshold onto that grid without changing
the reported instants. We therefore tune $\tau$, on unlabeled CSI only, so
that~\eqref{eq:rate} meets the target $\bar{\alpha}$. All CSI reporting rules above are thus compared at the same mean rate, and our threshold rule is rate-controlled, close in spirit to the
adaptive threshold of~\cite{varshney2023adaptive}.
%###
\vspace{-1 em}
\subsection{Spectral Resolution (Subcarrier Grouping)}\label{sec:reporting}
With grouping factor $g\in\{1,\dots,K\}$, each report carries the CSI of only
$N_{\mathsf{SC}}=\lfloor K/g\rfloor$ of the $K$ subcarriers,\footnote{Optimized subcarrier selection~\cite{behzad2025subcarrier,singh2025slimsense} could replace the fixed \(\mathbf{S}_g\); however, this is outside the scope of this study. Moreover, the additional computational complexity may not justify the resulting (limited) performance~gain.}
\begin{equation}
    \mathbf{Z}(t_m)=\tilde{\mathbf{X}}(t_m)\,\mathbf{S}_g\in\mathbb{C}^{N\times N_{\mathsf{SC}}},
    \label{eq:grouping}
\end{equation}
where $\mathbf{Z}(t_m)$ is the CSI report sent at instant $t_m$,
$\tilde{\mathbf{X}}(t_m)$ is the frame to be reported, given by~\eqref{eq:payload}, and
$\mathbf{S}_g\in\mathbb{R}^{K\times N_{\mathsf{SC}}}$ is a grouping matrix whose $j$-th
column maps the $j$-th group of $g$ adjacent subcarriers, i.e.,
$\mathcal{K}_j:=\{(j-1)g+1,\dots,jg\}$, to a single reported value. The last
$K-gN_{\mathsf{SC}}$ subcarriers, which belong to no group, are dropped. We consider two choices of $\mathbf{S}_g$ with the same report size:
\begin{itemize}
  \item \emph{Uniform decimation:} $[\mathbf{S}_g]_{k,j}=1$ if $k=(j-1)g+1$, i.e., the
  first subcarrier of $\mathcal{K}_j$, and $0$ otherwise, following the subcarrier
  grouping of 802.11bf~\cite[Sec.~9.4.1.81.3]{IEEE80211bf2025}.
  \item \emph{Block mean (proposed):} $[\mathbf{S}_g]_{k,j}=1/g$ if
  $k\in\mathcal{K}_j$, and $0$ otherwise, i.e., the average over the group.
\end{itemize}
Block mean is a moving-average low-pass filter applied before down-sampling. It changes
only the STA-side processing, not the report format, in line with the standard's note
that anti-aliasing may be needed before down-sampling at the largest grouping~\cite[Table~9-129h, NOTE~2]{IEEE80211bf2025}. For complex CSI datasets, it is
applied after removing the linear phase slope across subcarriers, so that timing offsets
do not attenuate the average. Since $g$ is defined relative to the $K$ subcarriers
available, which may already be grouped by the device, we sweep $g$ over a finer grid
than the standard's grouping values $\{4,8,16\}$ to trace the full trade-off.
%###
\vspace{-1 em}
\subsection{CSI Feedback Cost Model}
Following the split in~\cite[Eqs.~(17)--(19)]{varshney2023adaptive}, the
feedback cost has a part that does not depend on the CSI reporting rate and a
part that scales with the number of reports. The size of one CSI report
is~\cite[Eq.~(9-5e)]{IEEE80211bf2025}
\begin{align}
    c_{\mathsf{rep}}(N_{\mathsf{SC}}) = c_{\mathsf{h}}
    + 1.5N + 2N N_{\mathsf{SC}} + 2N_{\mathsf{R}}\, ,  (\text{in octets})
    \label{eq:report_size}
\end{align}
where $c_{\mathsf{h}}$ is the fixed frame overhead and "octet" is 8 bits. Threshold-based reporting introduces, at each instant, a trigger frame and a short report containing a 4-bit CSI-variation feedback field~\cite[Table~9-129i]{IEEE80211bf2025}; we denote this overhead by~\(c_{\mathsf{v}}\). The expected feedback cost per session is
\begin{align}
    \bar{C}(\pi,\bar{\alpha},g) =
    \mathds{1}\{\pi\in\Pi_{\mathsf{thr}}\}\,T c_{\mathsf{v}}
    + \bar{\alpha} T\, c_{\mathsf{rep}}\big(\lfloor K/g\rfloor\big),
    \label{eq:cost}
\end{align}
where \(\pi\) denotes the reporting rule and \(\Pi_{\mathsf{thr}}\) the set of threshold-based rules. Full-CSI reporting incurs
\(C_{\mathsf{full}}=T c_{\mathsf{rep}}(K)\).
Notably, the first term in~\eqref{eq:cost} is independent of \(\bar{\alpha}\): regardless of how rarely the channel crosses the threshold \(\tau\), threshold-based reporting incurs at least \(T c_{\mathsf{v}}\) in feedback overhead. This overhead is also present in~\cite{varshney2023adaptive}, but its impact on the accuracy--cost trade-off was not evaluated.
\\\indent 
CSI feedback cost/overhead is the budgeted resource in this work. The number of NDP soundings, \(S\), is a separate resource (and has its own cost): the STA must measure every sounding instance when performing temporal averaging or evaluating \(v(t)\). Table~\ref{tab:schemes} reports both resources for each scheme.
\begin{table}[t]
\small
\centering
\caption{Soundings and expected CSI feedback cost per session.}
\label{tab:schemes}
\setlength{\tabcolsep}{4pt}
\begin{tabular}{lll}
\hline
Scheme & Num. of~soundings & Feedback cost \\
\hline
Periodic, random          & $\bar{\alpha} T$ & $\bar{\alpha} T c_{\mathsf{rep}}$ \\
Periodic-mean             & $T$        & $\bar{\alpha} T c_{\mathsf{rep}}$ \\
Threshold, threshold-mean & $T$        & $T c_{\mathsf{v}}+\bar{\alpha} T c_{\mathsf{rep}}$ \\
Full CSI                  & $T$        & $T c_{\mathsf{rep}}(K)$ \\
\hline
\end{tabular}
\vspace{-1em}
\end{table}
\vspace{-1 em}
\subsection{Machine Learning Task and Pareto-frontier Reporting}
The AP reconstructs the CSI sequence over the measurement session using
zero-order hold (ZOH). Specifically, the most recently received CSI report
is retained until a subsequent report is available:
\begin{align}
    \hat{\mathbf{X}}(t)
    &= \mathbf{Z}(t_{m(t)}), \,
    m(t):=\max\{m:t_m\leq t\}, \,
    t=0,\ldots,T-1,
    \label{eq:zoh}
\end{align}
following the hold mechanism in~\cite[Eq.~(13)]{varshney2023adaptive}.
We assume that the first frame of each measurement window is always
reported, i.e., $t_1=0$, such that $\hat{\mathbf{X}}(t)$ is defined for
all ${t\in\{0,\ldots,T-1\}}$.

A CSI reporting scenario is defined as
%\begin{equation}
   $ \nu\,:=\,(\pi,\rho,\bar{\alpha},s,g),$
%\end{equation}
where $\pi\in\{\mathsf{Per},\mathsf{Rand},\mathsf{Thr}\}$ denotes respectively the periodic, random, and threshold-based reporting policy defined in Sec.~\ref{sec:temporal}, ${\rho\in\{\mathsf{inst},\mathsf{mean}\}}$ specifies the instant and mean payload type in~\eqref{eq:payload}, and ${\bar{\alpha}\in(0,1]}$ is the target
reporting rate defined in~\eqref{eq:rate}. Moreover, ${s\in\{\mathsf{Uni},\mathsf{BM}\}}$
denotes the uniform decimation and block mean subcarrier grouping scheme in Sec.~\ref{sec:reporting}, and ${g\in\mathcal{G}\subseteq\{1,\ldots,K\}}$
is the grouping factor defined in~\eqref{eq:grouping}. Let
$\mathcal{V}$ denote the set of all reporting scenarios.

Each scenario $\nu$ represents a \textit{fixed} configuration of the CSI reporting mechanism. For the threshold-based reporting, i.e., $\pi=\mathsf{Thr}$, the threshold $\tau$ is determined by a bisection search such that the resulting average reporting rate satisfies
$\mathbb{E}[M]/T=\bar{\alpha}$. Let $\hat{\mathbf{X}}_{\nu}(t)$ denote the corresponding reconstructed CSI instance $t$ at the AP.

For \textit{each} $\nu\in\mathcal{V}$, the AP performs session-level classification
from the reconstructed CSI sequence
\begin{equation}
    \hat{\mathcal{X}}_{\nu}
    :=\{\hat{\mathbf{X}}_{\nu}(t)\}_{t=0}^{T-1}
\end{equation}
using a neural network (NN) $f_{\boldsymbol{\theta}}$ parameterized by $\boldsymbol{\theta}$. A separate NN is
trained for each reporting scenario, with its parameters optimized by
minimizing the empirical cross-entropy loss:
\begin{align}
    \boldsymbol{\theta}^{\star}(\nu)
    =
    \arg\min_{\boldsymbol{\theta}}
    \frac{1}{|\mathcal{D}_{\mathsf{tr}}|}
    \sum_{(\mathcal{X},y)\in\mathcal{D}_{\mathsf{tr}}}
    \ell\!\left(
        f_{\boldsymbol{\theta}}
        \big(\hat{\mathcal{X}}_{\nu}(\mathcal{X})\big),
        y
    \right),
    \label{eq:training}
\end{align}
where $\mathcal{D}_{\mathsf{tr}}$ denotes the training dataset and
$\ell(\cdot,\cdot)$ is the cross-entropy loss. The resulting model is
evaluated on sessions excluded from $\mathcal{D}_{\mathsf{tr}}$. The
corresponding (recognition) \textit{accuracy} is defined as
\begin{align}
    A(\nu)
    =
    \Pr\!\left[
        f_{\boldsymbol{\theta}^{\star}(\nu)}
        \big(\hat{\mathcal{X}}_{\nu}\big)
        =y
    \right],
    \label{eq:accuracy}
\end{align}
where the probability is with respect to the joint distribution of the CSI session and its activity label, i.e., $(\mathcal X,y)$ and, for $\pi=\mathsf{Rand}$, also the random reporting mask.

Each scenario is also associated with an average feedback cost
$\bar{C}(\nu)$ as defined in~\eqref{eq:cost}. The feedback cost depends
only on $(\pi,\bar{\alpha},g)$, since neither the payload type $\rho$ nor the
grouping scheme $s$ changes the number of reported CSI values. Consequently,
scenarios that differ only in $\rho$ or $s$ have identical feedback costs
and can be compared directly in terms of their end task accuracy.

The evaluated scenarios are compared using the accuracy--cost Pareto
frontier. A scenario $\nu\in\mathcal{V}$ is Pareto-efficient if there is no
$\nu'\in\mathcal{V}$ such that
\begin{equation} \label{eq:pareto}
    \bar{C}(\nu')\leq\bar{C}(\nu)
    \quad\text{and}\quad
    A(\nu')>A(\nu).
\end{equation}
For a given cost level, the Pareto-efficient set in~\eqref{eq:pareto} thus gives the highest accuracy attained within $\mathcal{V}$. By~\eqref{eq:cost}, this cost factors into a temporal and a spectral part,
\begin{align}
\frac{\bar C}{C_{\mathsf{full}}}
 = \frac{M}{T}\cdot\frac{c_{\mathsf{rep}}(N_{\mathsf{SC}})}{c_{\mathsf{rep}}(K)}
 + \mathds{1}\{\pi\in\Pi_{\mathsf{thr}}\}\,\frac{c_\mathsf{v}}{c_{\mathsf{rep}}(K)},
\label{eq:split}
\end{align}
where the last term is the minimum cost of threshold reporting. For each dataset, Sec.~\ref{sec:sim} reports the Pareto frontier of each scheme and how its cost splits between the two factors in~\eqref{eq:split}.
%###
\vspace{-1 em}
\section{Neural Network Model and Training}\label{sec:method}
\textit{Neural Network Input:} The ZOH in~\eqref{eq:zoh} is causal (the first frame is always reported, so
every frame is defined from past CSI reports only) and keeps the input length at
$T$ for every $(\pi,\bar{\alpha})$, so one architecture serves all settings and
needs no padding or masking. Furthermore, the report timing is implicit in
the input, as the start and length of each held segment. The reconstruction
is applied at native subcarrier resolution, after which subcarrier grouping
yields $N_{\mathsf{SC}}$ subcarriers and the result is normalized. Each input
$\hat{\mathcal{X}}_{\nu}$ then enters the network as a frequency--time map
with $W$ channels, $N_{\mathsf{SC}}$ rows and $T$ columns, where $W=2N$
(amplitude and phase of the $N$ receive antennas as separate channels) when
both CSI amplitude and phase are available, and $W=N$ for amplitude-only
datasets. When used, the phase is sanitized and then held and grouped
exactly like the amplitude. 
% The dataset-specific values of $N$,
% $N_{\mathsf{SC}}$, $T$ and the normalization are given in
% Sec.~\ref{sec:sim}.

\begin{figure}[t]
%\centering
\hspace{-2 em}
\resizebox {\columnwidth}{!}{\usetikzlibrary{arrows.meta,positioning,calc,decorations.pathreplacing}
\definecolor{inC}{HTML}{378ADD}\definecolor{c1}{HTML}{534AB7}\definecolor{c2}{HTML}{D85A30}
\definecolor{pl}{HTML}{BA7517}\definecolor{gr}{HTML}{1D9E75}\definecolor{hd}{HTML}{D4537E}
\definecolor{txt}{HTML}{3A3A38}
\tikzset{
  font=\scriptsize, text=txt,
  arr/.style={-{Stealth[length=3pt,width=3pt]},line width=0.5pt,draw=txt},
  stage/.style={rounded corners=2pt,draw=#1!70,fill=#1!6,line width=0.4pt,anchor=north west},
  ttl/.style={font=\scriptsize\bfseries,text=#1!80!black,inner sep=1.5pt},
  dim/.style={font=\tiny,text=txt,inner sep=1pt},
  op/.style={rounded corners=1pt,fill=#1!20,draw=#1!60,line width=0.3pt,inner sep=1.2pt,font=\tiny,align=center},
}
% cuboid: x, y (front bottom-left), w, h, depth, color
\newcommand{\cub}[6]{%
  \fill[#6!35,draw=#6!80!black,line width=0.3pt] (#1+#5,#2+#5) rectangle ++(#3,#4);
  \foreach \k in {0.33,0.66}{\draw[#6!80!black,line width=0.2pt,fill=#6!30] ($(#1,#2)+(\k*#5,\k*#5)$) rectangle ++(#3,#4);}
  \fill[#6!22,draw=#6!80!black,line width=0.3pt] (#1,#2) rectangle ++(#3,#4);
  \draw[#6!80!black,line width=0.3pt] (#1,#2+#4)--(#1+#5,#2+#4+#5) (#1+#3,#2+#4)--(#1+#3+#5,#2+#4+#5) (#1+#3,#2)--(#1+#3+#5,#2+#5);
}

\begin{tikzpicture}
\def\H{2.45}
%================ ROW 1 =================
\node[stage=inC,minimum width=1.75cm,minimum height=\H cm] (sIn) at (0,0) {};
\node[ttl=inC,anchor=north] at (sIn.north) {Input CSI};
\begin{scope}[shift={(0.52,-1.62)}]
  \foreach \k in {2,1,0}{
    \begin{scope}[shift={(\k*0.11,\k*0.11)}]
      \fill[inC!12,draw=inC!80!black,line width=0.3pt] (0,0) rectangle (0.84,0.98);
      \ifnum\k=0
        \foreach \i in {0,...,6}{\foreach \j in {0,...,6}{
          \pgfmathsetmacro{\v}{int(max(8,min(90,45+40*sin(55*\i+30*\j)*cos(35*\j-20*\i))))}
          \fill[inC!\v!white] (\i*0.12,\j*0.14) rectangle ++(0.12,0.14);}}
        \draw[inC!80!black,line width=0.3pt] (0,0) rectangle (0.84,0.98);
      \fi
    \end{scope}}
  \draw[|-|,line width=0.3pt] (0,-0.08) -- node[below,dim]{$T$} (0.84,-0.08);
  \draw[|-|,line width=0.3pt] (-0.08,0) -- node[left,dim]{$N_{\mathsf{SC}}$} (-0.08,0.98);
  \node[dim] at (1.02,1.25) {$W$};
\end{scope}
\node[dim,anchor=south,align=center] at ($(sIn.south)+(0,0.04)$) {amplitude (+ phase)\\of Rx antennas};

% --- CNN
\node[stage=c1,minimum width=3.85cm,minimum height=\H cm] (sCnn) at (1.9,0) {};
\node[ttl=c1,anchor=north] at (sCnn.north) {2D CNN feature extractor};
\node[dim,anchor=north] at (2.85,-0.32) {\textbf{Block 1:} $5{\times}5$, $W{\to}32$};
\node[dim,anchor=north] at (4.80,-0.32) {\textbf{Block 2:} $3{\times}3$, $32{\to}64$};
\cub{2.45}{-1.46}{0.58}{0.62}{0.20}{c1}
\cub{4.48}{-1.36}{0.40}{0.46}{0.26}{c2}
\draw[arr] (3.30,-1.12) -- (4.38,-1.12);
\node[dim,anchor=north] at (2.85,-1.50) {$32{\times}\frac{T}{2}{\times}\frac{N_{\mathsf{SC}}}{2}$};
\node[dim,anchor=north] at (4.80,-1.50) {$64{\times}T'{\times}\frac{N_{\mathsf{SC}}}{4}$};
\node[op=c1,anchor=south,text width=3.55cm] at ($(sCnn.south)+(0,0.06)$) {each block: Conv, BN, ReLU,\\spatial dropout, $2{\times}2$ max pool};

% --- Adaptive pool
\node[stage=pl,minimum width=1.93cm,minimum height=\H cm] (sPl) at (5.9,0) {};
\node[ttl=pl,anchor=north,align=center] at (sPl.north) {Adaptive pool};
\node[dim,anchor=north] at ($(sPl.north)+(0,-0.26)$) {avg.\ over frequency};
\cub{6.52}{-1.30}{0.40}{0.46}{0.26}{pl}
\draw[decorate,decoration={brace,amplitude=2pt,mirror},line width=0.3pt] (6.52,-1.35)--node[below,dim,yshift=-1pt]{$D$ bands}(6.92,-1.35);
\node[dim,align=center,anchor=south] at ($(sPl.south)+(0,0.04)$) {$64{\times}T'{\times}D$,\\$T'\approx T/4$, any $N_{\mathsf{SC}}$};

\draw[arr] (sIn.east) -- (sCnn.west);
\draw[arr] (sCnn.east) -- (sPl.west);

%================ ROW 2 =================
\def\yy{-2.72}
\node[stage=gr,minimum width=3.65cm,minimum height=\H cm] (sGru) at (0,\yy) {};
\node[ttl=gr,anchor=north] at (sGru.north) {Bidirectional GRU};
\node[dim,anchor=north] at ($(sGru.north)+(0,-0.26)$) {$L$ layers, hidden size $H$};
\foreach \i/\lab in {0/1,1/2,3/{T'}}{
  \pgfmathsetmacro{\xx}{0.55+0.76*\i}
  \fill[pl!30,draw=pl!80!black,line width=0.3pt] (\xx-0.06,\yy-2.28) rectangle ++(0.12,0.34);
  \node[dim,anchor=east] at (\xx-0.06,\yy-2.11) {$\mathbf{x}_{\lab}$};
  \node[draw=gr!80!black,fill=gr!15,rounded corners=1pt,minimum width=0.46cm,minimum height=0.56cm,line width=0.3pt] (g\i) at (\xx,\yy-1.30) {};
  \node[dim] at (\xx,\yy-1.17) {$\overrightarrow{h}$};
  \node[dim] at (\xx,\yy-1.43) {$\overleftarrow{h}$};
  \draw[arr] (\xx,\yy-1.92) -- (g\i.south);
  \draw[arr] (g\i.north) -- (\xx,\yy-0.72);
}
\node at (0.55+0.76*2,\yy-1.30) {$\cdots$};
\foreach \a/\b in {0/1}{
  \draw[arr,gr!70!black] ($(g\a.east)+(0,0.10)$) -- ($(g\b.west)+(0,0.10)$);
  \draw[arr,gr!70!black] ($(g\b.west)+(0,-0.10)$) -- ($(g\a.east)+(0,-0.10)$);}
\draw[arr,gr!70!black] ($(g1.east)+(0,0.10)$) -- ++(0.22,0);
\draw[arr,gr!70!black] ($(g1.east)+(0.22,-0.10)$) -- ($(g1.east)+(0,-0.10)$);
\draw[arr,gr!70!black] ($(g3.west)+(-0.22,0.10)$) -- ($(g3.west)+(0,0.10)$);
\draw[arr,gr!70!black] ($(g3.west)+(0,-0.10)$) -- ++(-0.22,0);
\node[dim,anchor=west] at (2.87,\yy-2.11) {$\in\!\mathbb{R}^{64D}$};
\node[dim,anchor=west] at (2.87,\yy-0.82) {$\in\!\mathbb{R}^{2H}$};

% --- Temporal pyramid
\node[stage=c1,minimum width=2.65cm,minimum height=\H cm] (sTp) at (3.78,\yy) {};
\node[ttl=c1,anchor=north] at (sTp.north) {Temporal pyramid};
\node[dim,anchor=north] at ($(sTp.north)+(0,-0.26)$) {segment means over time};
\def\px{4.33}\def\pw{1.55}
\foreach \r/\n in {0/1,1/2,2/4}{
  \pgfmathsetmacro{\py}{\yy-0.78-0.36*\r}
  \node[dim,anchor=east] at (\px-0.03,\py-0.11) {\n};
  \pgfmathsetmacro{\sw}{\pw/\n}
  \foreach \s in {1,...,\n}{
    \fill[c1!22,draw=c1!80!black,line width=0.3pt] (\px+\sw*\s-\sw+0.02,\py-0.22) rectangle ++(\sw-0.04,0.22);}
}
\node[dim,anchor=south,align=center] at ($(sTp.south)+(0,0.04)$) {concat.:\ $(1{+}2{+}4)\,2H=14H$};

% --- Classifier
\node[stage=hd,minimum width=1.38cm,minimum height=\H cm] (sHd) at (6.59,\yy) {};
\node[ttl=hd,anchor=north] at (sHd.north) {Classifier};
\node[op=hd,minimum width=1.05cm] (dp) at (7.14,\yy-0.58) {Dropout};
\node[op=hd,minimum width=1.05cm] (ln) at (7.14,\yy-0.98) {Linear};
\draw[arr] (dp) -- (ln);
\foreach \i/\h in {0/0.16,1/0.40,2/0.09,3/0.24}{
  \fill[hd!30,draw=hd!80!black,line width=0.3pt] (6.79+0.19*\i,\yy-1.98) rectangle ++(0.13,\h);}
\draw[arr] (ln) -- (7.14,\yy-1.52);
\node[dim,anchor=south,align=center] at ($(sHd.south)+(0,0.04)$) {class/activity\\scores};

\draw[arr] (sGru.east) -- (sTp.west);
\draw[arr] (sTp.east) -- (sHd.west);
% reshape connector
\draw[arr,rounded corners=2pt] (sPl.south) -- ++(0,-0.08) -| (sGru.north);
\end{tikzpicture}
%\end{document}
}
\caption{Proposed CNN-GRU with temporal pyramid neural network~architecture.}
\label{fig:cnn_gru}
\vspace{-1 em}
\end{figure}
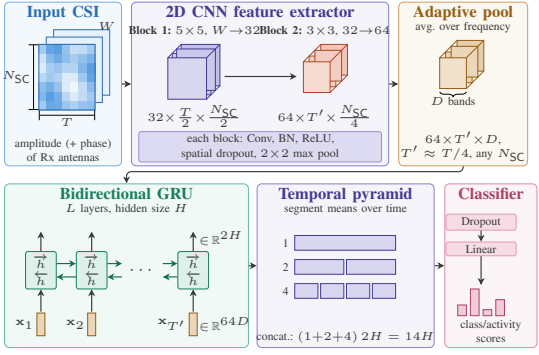

\textit{Neural Network Architecture:} We use a  convolutional neural network
combined with a gated recurrent unit (CNN-GRU) with a temporal pyramid architecture, as shown in~Fig.~\ref{fig:cnn_gru}. The input is a $W\times T\times N_{\mathsf{SC}}$
tensor, whose $W$ channels hold the CSI amplitude (and, where available, the
sanitized phase) of the receive antennas over $T$ time steps and
$N_{\mathsf{SC}}$ subcarriers. Two convolutional blocks ($5\times5$ and
$3\times3$ kernels, $W\!\to\!32\!\to\!64$ channels, each with batch
normalization, ReLU, spatial dropout and $2\times2$ max pooling) extract
local time--frequency features and shorten time to $T'\approx T/4$ steps.
An adaptive average pool then collapses the frequency axis to $D$ bands, so
that for any $N_{\mathsf{SC}}$, i.e., for any subcarrier grouping, each time
step is described by a $64D$-dimensional feature. A bidirectional GRU with
$L$ layers and hidden size $H$ models the temporal dynamics. A temporal
pyramid readout averages the $2H$-dimensional GRU outputs over $1$, $2$ and
$4$ equal time segments and concatenates them into a $14H$-dimensional
vector, which keeps the coarse temporal order of the activity. Because only
the time axis varies with the input and the readout has a fixed size, the
same network accepts any window length $T$ and any subcarrier grouping.
Finally, a classifier head (dropout followed by a linear layer) gives the
class scores. 
%The dataset-specific values of $C$, $D$, $L$, $H$, the dropout rates and the training settings are given in~Sec.~\ref{sec:sim}.
\\\indent 
The backbone and the temporal-pyramid readout were selected in a preliminary
comparison on SignFi against CNN-only, attention- and statistics-pooling, and
Transformer-based alternatives, whereas the layer sizes follow common CNN--GRU
practice, with $(D,H,L)$ scaled to each dataset (Table~\ref{tab:setup}). The
architecture is fixed across all $(\bar{\alpha},g)$ settings, so that accuracy differences
reflect the reporting configuration rather than model tuning.

\textit{Training procedure:} For any given $(\pi,\bar{\alpha},g)$ and dataset, we train the neural network by the following steps:
\\ 1) \emph{Report masks.} Reporting instants are generated for each session
        according to rule $\pi$ at rate $\bar{\alpha}$. For threshold-based
        rules, $\tau$ is calibrated independently for each responder using
        unlabeled CSI to match the rate $\bar{\alpha}$. For the random
        rule, a new mask is generated for each session in each Monte Carlo~run.
\\ 2) \emph{Inputs.} Training and test sessions are reconstructed using
        ZOH~\eqref{eq:zoh} under their respective reporting masks, then grouped
        with factor $g$ and normalized.
\\ 3) \emph{Optimization.} A validation subset is held out from the
        training sessions. The network parameters $\boldsymbol{\theta}$ are
        randomly initialized and optimized using cross-entropy loss, with
        linear learning-rate warm-up followed by cosine decay. The checkpoint
        with the best validation performance is retained.
\\ 4)  \emph{Testing/inference.} The trained model is evaluated on the \textit{held-out} test sessions. Accuracy is averaged across Monte Carlo runs and evaluation
        folds. Simulation parameters are given
in Sec.~\ref{sec:sim}.

\vspace{-1 em}
\section{Simulation Results}\label{sec:sim}
This section evaluates the accuracy--cost tradeoffs of the CSI reporting schemes. We first describe the datasets and simulation setup, and then discuss the results.
\vspace{-1 em}
\subsection{Datasets and Simulation Setup}
We emulate the reporting schemes on three public CSI datasets that differ in link size, bandwidth, and task (Table~\ref{tab:setup}), treating each session as $T\!=\!200$ sounding instances. SignFi~\cite{ma2018signfi} contains $1500$ sessions for each of five users, each with exactly $T$ instances. NTU-Fi HAR~\cite{yang2023sensefi} has a fixed split of $936$ training and $264$ test sessions without user labels; each $2000$-sample session is decimated by $4$. CSI-HAR~\cite{moshiri2021csi} contains $20$ sessions per activity for each of three subjects, of which we use the first $500$ instances. For NTU-Fi HAR and CSI-HAR, we take the central $T$ instances of the resulting $500$-instance sequence.
\\\indent 
Each scheme in Sec.~\ref{sec:model} is applied to the full-resolution CSI of every session. For the threshold rule, $\tau$ is found by bisection so that the mean reporting rate equals $\bar\alpha$, using unlabeled CSI only: per user for SignFi, per subject for CSI-HAR, and separately on the training and test sets for NTU-Fi HAR, which has no user labels.\footnote{Because NTU-Fi HAR and CSI-HAR provide only CSI amplitude, we evaluate~\eqref{eq:variation} on $|\mathbf{X}(t)|$ for all datasets. On SignFi, evaluating it on the complex CSI after removing the linear phase slope improves threshold reporting by up to $7.6$ points at $g\!=\!1$, but it remains below periodic reporting at every rate.} The feedback cost follows~\eqref{eq:cost} with $c_{\mathsf h}=28$ and $c_{\mathsf v}=29$ octets, where $\bar\alpha T$ is replaced by the realized number of reports $M$, averaged over the test sessions, and we report the relative cost $\bar C/C_{\mathsf{full}}$ given in~\eqref{eq:split}.
\\\indent 
The CNN-GRU of Sec.~\ref{sec:method} is trained from scratch for every scenario $\nu$ with the settings in Table~\ref{tab:setup}. SignFi and CSI-HAR are evaluated leave-one-user-out ($5$ folds) and leave-one-subject-out ($3$ folds), respectively, since random splits inflate accuracy through static per-session signatures (from ${\sim}61\%$ to ${\sim}93\%$ on CSI-HAR~\cite{varga2025partition}); NTU-Fi HAR uses its fixed split. Each fold is trained over $50$ Monte Carlo runs, each with its own validation split and initialization. Run~$i$ uses the same split and seed in every scenario, so comparisons between scenarios are paired. We report the mean of the per-fold accuracies and its standard error, taken over folds, or over runs for NTU-Fi HAR.\footnote{With both grouping choices, random reporting was below periodic reporting at the same mean rate by $13.0$, $2.3$, and $0.5$ percentage points on SignFi, CSI-HAR, and NTU-Fi HAR, respectively, and showed the same grouping trends; it is omitted for brevity.}

\begin{table}[t]
\centering
\footnotesize
\setlength{\tabcolsep}{3pt}
\caption{Datasets and simulation parameters ($T=200$).}
\label{tab:setup}
\begin{tabular}{llll}
\hline
 & SignFi & NTU-Fi HAR & CSI-HAR \\
\hline
Band & 20\,MHz & 40\,MHz & 20\,MHz \\
Link $N_{\mathsf T}\times N_{\mathsf R}$ & $1\times3$ & $1\times3$ & $1\times1$ \\
Subcarriers $K$ & 30 & 114 & 52 \\
Classes & 150 & 6 & 7 \\
Sessions & 7500 & 936\,/\,264 & 420 \\
Input channels $W$ & 6 (amp.+phase) & 3 (amp.) & 1 (amp.) \\
Evaluation split & LOUO (5 folds) & fixed train/test & LOSO (3 folds) \\
$\tau$ calibrated per & user & split & subject \\
Std.\ error over & folds & MC runs & folds \\
Rate $\bar\alpha$ & \multicolumn{3}{c}{$\{1,\,1/2,\,1/4,\,1/8,\,1/16\}$} \\
Grouping $g$ & $\{1,2,3,5,6,10\}$ & $\{1,2,4,8,16\}$ & $\{1,2,4,8,13\}$ \\
Cost $c_{\mathsf h}$\,/\,$c_{\mathsf v}$ (octets) & \multicolumn{3}{c}{28\,/\,29} \\
CNN-GRU $(D,H,L)$ & $(8,128,2)$ & $(8,128,2)$ & $(4,64,1)$ \\
%Parameters & 1.08\,M & 0.82\,M & 0.15\,M \\
Optimizer & AdamW & Adam & Adam \\
LR\,/\,weight decay & $10^{-3}$\,/\,$10^{-4}$ & $10^{-3}$\,/\,$3{\cdot}10^{-4}$ & $10^{-3}$\,/\,$3{\cdot}10^{-4}$ \\
Epochs (warm-up) & 80 (3) & 100 (5) & 100 (5) \\
Batch size & 128 & 64 & 64 \\
Validation split & 10\,\% & 20\,\% & 20\,\% \\
Checkpoint & val.\ acc. & val.\ acc. & val.\ loss \\
Normalization & centering & z-score & z-score \\
Monte Carlo runs & 50 & 50 & 50 \\
\hline
\end{tabular}
\vspace{-1em}
\end{table}
\vspace{-1 em}
\subsection{Performance Evaluations and Discussions}
\begin{table}[t]
%\footnotesize
\caption{Highest accuracy (\%) achieved within each relative feedback cost budget $\bar C/C_{\mathsf{full}}$. The best result per budget is bold; dashes (--) mark settings below the minimum threshold cost.}
\label{tab:ranking}
\centering
\setlength{\tabcolsep}{4pt}
\begin{tabular}{llccc}
\hline
Dataset & CSI Reporting scheme & $2\%$ & $10\%$ & $25\%$ \\
\hline
SignFi      & Per(mean)+BM  & \textbf{57.5} & \textbf{61.5} & \textbf{63.1} \\ % (full CSI 62.6)
 & Per(inst)+Uni & 34.5 & 48.4 & 53.6 \\
            & Thr(mean)+BM  & --   & --   & 36.9 \\
            & Thr(inst)+Uni & --   & --   & 31.5 \\
\hline
CSI-HAR     & Per(inst)+Uni & \textbf{40.3} & \textbf{49.9} & \textbf{51.3} \\ % (full CSI 49.5)
 & Per(mean)+BM  & 39.3 & 47.3 & 49.9 \\
            & Thr(inst)+Uni & --   & --   & 40.9 \\
            & Thr(mean)+BM  & --   & --   & 37.1 \\
\hline
NTU-Fi      & Per(mean)+BM  & \textbf{97.2} & 97.8 & 98.1 \\ % (full CSI 98.6)
 & Per(inst)+Uni & 96.7 & 97.7 & 98.2 \\
            & Thr(inst)+Uni & --   & \textbf{97.9} & 98.5 \\
            & Thr(mean)+BM  & --   & 97.5 & \textbf{98.6} \\
\hline
\end{tabular}
\end{table} 

First, Table~\ref{tab:ranking} summarizes Pareto-efficient accuracy performance for different \textit{relative} CSI feedback cost ratios. The optimal scheme varies across different budgets and datasets for three key reasons. (i) Threshold reporting requires a minimum baseline cost, $c_\mathsf{v}/c_{\mathsf{rep}}(K)$ (13.3\%, 21.4\%, and 4.0\% for the respective datasets). Below these limits, threshold schemes are infeasible, leaving only periodic schemes viable. (ii) Utilizing a mean payload is advantageous for tasks that rely on fine, rapid temporal details. For instance, on the SignFi dataset (150 signs), it boosts accuracy by 23 percentage points at a strict 2\% budget. Conversely, on CSI-HAR (7 coarse activities, single antenna), it offers no gain beyond one standard error, making Per(inst)+Uni the most effective choice. (iii) As the budget increases, the accuracy of all schemes converges toward that of full CSI. On NTU-Fi, for budgets above 10\%, the performance gap narrows to less than 0.5 percentage points, making all schemes equally viable.
\begin{figure}
    \centering
    \includegraphics[width=0.65\linewidth]{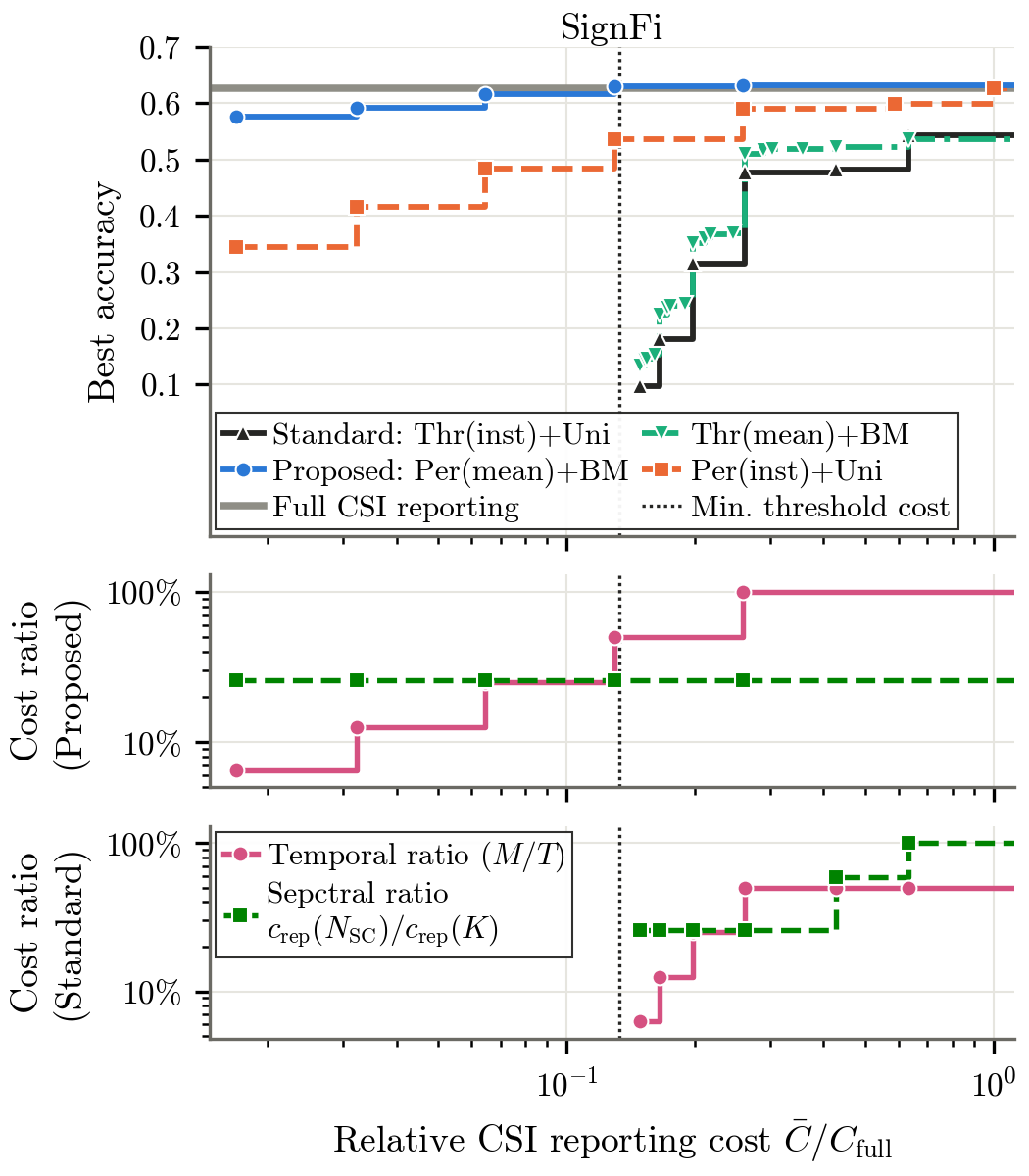}
    \caption{
    \small
    Accuracy–cost Pareto frontiers on SignFi (top) and temporal report ratio and spectral size ratio along the proposed and standard schemes (bottom).
    \textbf{Legend}: ``Thr" and ``Per" denote threshold and periodic reporting; ``inst" and ``mean" denote the instant and mean payloads in~\eqref{eq:payload}; ``Uni" and ``BM" denote uniform and block-mean subcarrier grouping in~\eqref{eq:grouping}. Standard is Thr(inst)+Uni (IEEE
802.11bf) and Proposed is Per(mean)+BM. Full CSI reporting corresponds to $\bar{\alpha}=1$,
$g=1$, and ``Min.\ threshold cost" is $c_{\mathsf v}/c_{\mathsf{rep}}(K)$, the lowest cost at which threshold reporting can~operate.
    }
    \label{fig:pareto_signfi}
    \vspace{-1 em}
\end{figure}
\\\indent 
Fig.~\ref{fig:pareto_signfi} shows accuracy--cost frontiers for the SignFi dataset. Each curve gives the best accuracy attainable at or below relative cost $\bar C/C_{\mathsf{full}}$, following~\eqref{eq:pareto}; markers denote Pareto-efficient $(\bar{\alpha},g)$ settings. Accuracies average five held-out users and 50 Monte Carlo runs. The dotted line marks the minimum threshold cost, $c_{\mathsf v}/c_{\mathsf{rep}}(K)=13.3\%$. The lower panels decompose cost along the proposed and standard frontiers into report ratio $M/T$ and size ratio $c_{\mathsf{rep}}(N_{\mathsf{SC}})/c_{\mathsf{rep}}(K)$; their product gives relative cost, with $c_{\mathsf v}/c_{\mathsf{rep}}(K)$ added for threshold reporting. The proposed scheme reaches $57.5\%$ accuracy at $1.7\%$ of full cost, $61.5\%$ at $6.5\%$, and $63.1\%$ at $12.9\%$, matching full CSI ($62.6\%$). At the same costs, periodic reporting with instant payload and uniform grouping achieves $34.5\%$, $48.4\%$, and $53.6\%$, respectively, indicating that the gain comes from report formation rather than schedule. The standard threshold scheme reaches $9.7\%$ just above its minimum cost and $54.3\%$ at $63\%$ of full cost; mean payload with block-mean grouping raises this only to $13.4\%$ at $14.9\%$. Along the proposed frontier, grouping remains at $g=10$ ($N_{\mathsf{SC}}=3$, size ratio $25.9\%$), while the report ratio rises from $6.5\%$ to $100\%$. Thus, block-mean grouping allows aggressive spectral reduction, while added budget is best spent on temporal resolution. The standard frontier increases both ratios and remains below the proposed one. Standard errors are at most $0.1$, and paired per-fold differences have the same sign across all five folds.
\begin{figure}
    \centering
    \includegraphics[width=0.65\linewidth]{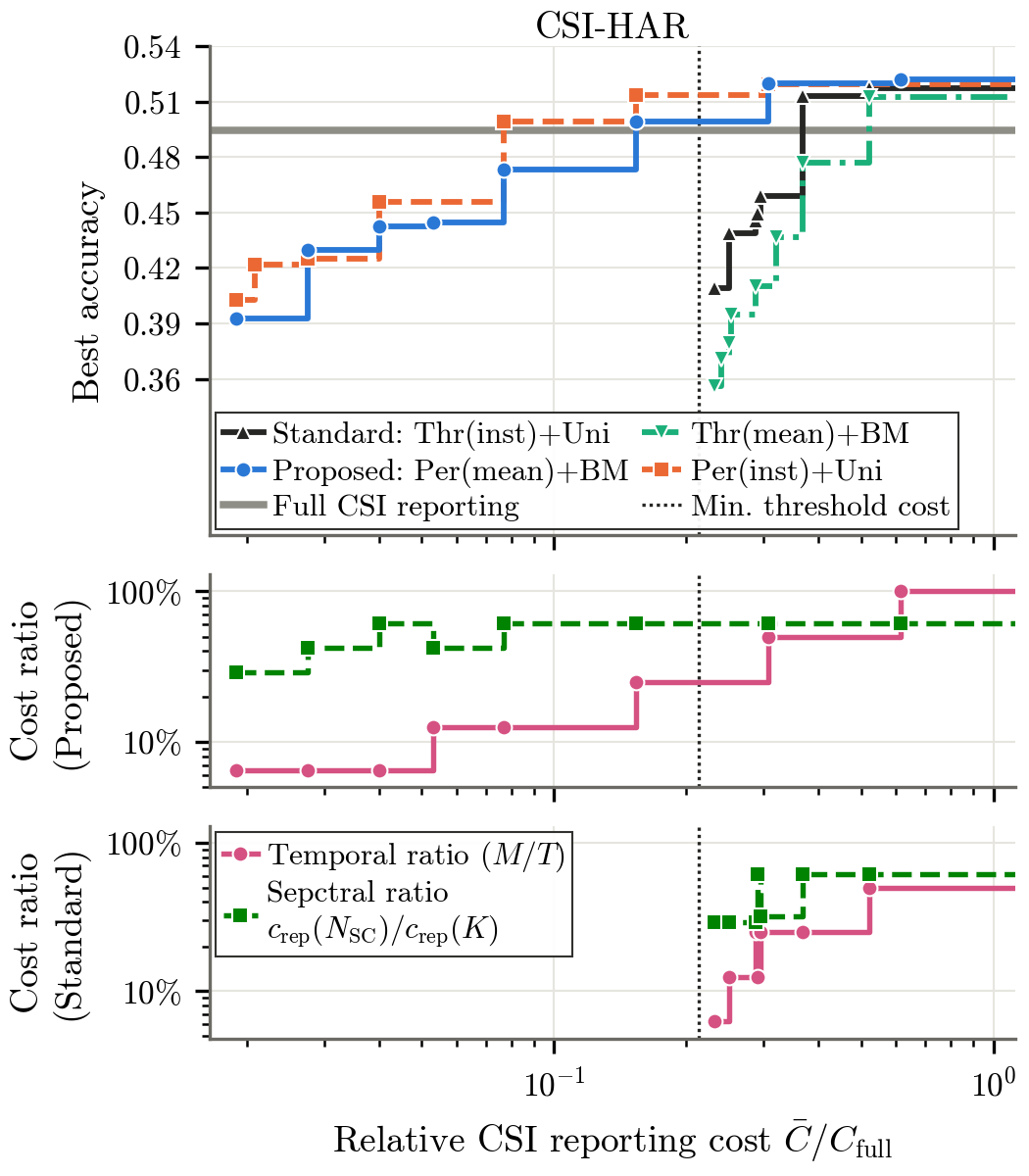}
    \caption{Accuracy–cost Pareto frontiers on CSI-HAR (top) and temporal report ratio and spectral size ratio along the proposed and standard schemes (bottom). Legend reading is the same as in~Fig.~\ref{fig:pareto_signfi}.}
    \label{fig:pareto_csihar}
    \vspace{-1 em}
\end{figure}
\\\indent 
Fig.~\ref{fig:pareto_csihar} shows accuracy--cost frontiers for the CSI-HAR dataset, where each accuracy is the average over three held-out subjects and 50 Monte Carlo runs. This single-antenna $20$\,MHz link has the highest minimum threshold cost ($21.4\%$), as a full report carries only $K=52$ subcarriers. Full-CSI accuracy is $49.5\%$, and per-setting standard errors (up to $0.05$) are comparable to differences between schemes. Periodic reporting reaches $40$--$42\%$ at $1.9$--$2.1\%$ of full cost and exceeds full-CSI accuracy ($\geq 49.9\%$) from $7.7\%$ onward. Threshold reporting reaches $40.9\%$ at its minimum cost and requires $36.8\%$ of full cost to reach $51.3\%$. Mean payload provides no measurable gain: the proposed and periodic instant-payload frontiers remain within one standard error. The proposed frontier moves early to moderate grouping ($g=2$--$4$, or $13$--$26$ subcarriers; size ratio $42$--$62\%$), reaches $g=2$ at $4\%$ cost, and then increases only the report ratio. 
%Grouping beyond $g=4$ removes useful information the classifier needs, so moderate spectral resolution is worth its cost. 
%Although some settings exceed full-CSI accuracy (e.g., $52.2\%$ at $g=2$), selecting the best of many noisy settings biases the frontier upward by a comparable amount; these results therefore do not establish a benefit from grouping.
\begin{figure}
    \centering
    \includegraphics[width=0.65\linewidth]{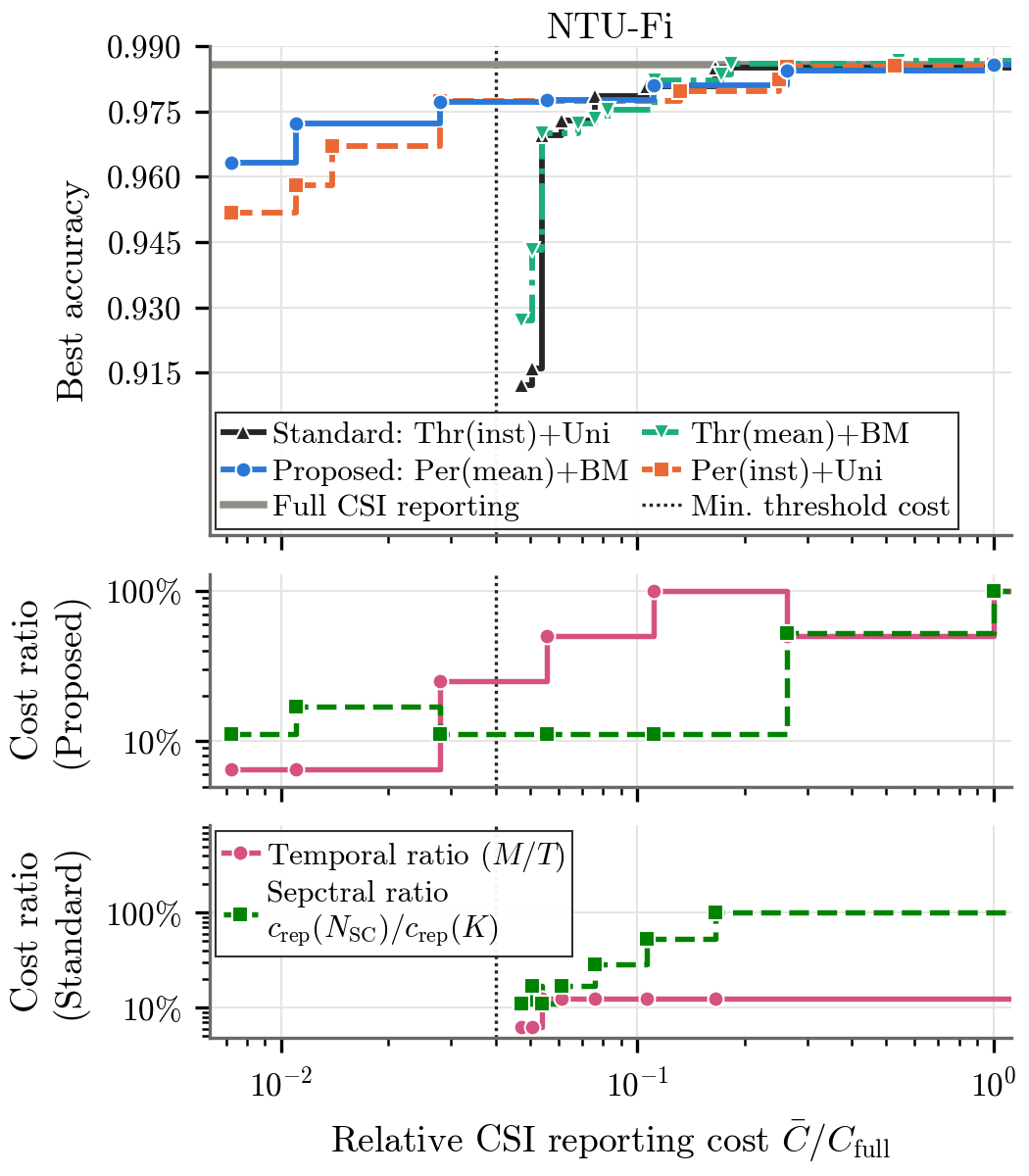}
    \caption{Accuracy–cost Pareto frontiers on the NTU-Fi dataset (top) and temporal report ratio and spectral size ratio along the proposed and standard schemes (bottom).}
    \label{fig:pareto_ntufi}
    \vspace{-1.5 em}
\end{figure}
\\\indent 
Fig.~\ref{fig:pareto_ntufi} shows accuracy--cost frontiers for the NTU-Fi HAR dataset. With $K=114$ subcarriers per report, the minimum threshold relative cost is $4.0\%$. All schemes approach full-CSI accuracy ($98.6\%$). At low cost, the proposed scheme reaches $96.3\%$ at $0.7\%$ cost and $97.2\%$ at $1.1\%$, compared with $95.2\%$ and $95.8\%$ for periodic instant-payload reporting with uniform grouping. Near its minimum cost, the standard scheme reaches $91.2\%$ at $4.7\%$ and $91.6\%$ at $5.1\%$; it requires $5.4\%$ cost to reach $97.0\%$, which the proposed scheme reaches at $2.8\%$. Above about $10\%$ cost, the frontiers converge near full-CSI accuracy. The proposed frontier keeps the size ratio at $11$--$17\%$ ($g=16$ or $8$, i.e., $7$--$14$ subcarriers) while increasing the report ratio from $6.5\%$ to $100\%$; finer grouping appears only near full cost. By contrast, the standard frontier fixes the report ratio at $12.5\%$ and spends additional budget on subcarriers, reducing accuracy at low budgets.

\vspace{-1 em}
\section{Conclusions}\label{sec_con}
We studied joint temporal and spectral cost on CSI reporting in Wi-Fi ISAC with HAR applications.
The results showed that CSI feedback could be substantially reduced without a proportional loss in HAR accuracy. Across the evaluated datasets, temporal resolution provided greater accuracy benefit per unit of feedback cost than spectral resolution, making increased reporting frequency with coarse subcarrier grouping more effective than retaining dense spectral information. 
Averaging CSI measurements between reports further improved the use of the available feedback budget. Consequently, full CSI was not Pareto-efficient in the evaluated settings: configurations using at most $7.7\%$ of the full-CSI feedback cost achieved $95.7$--$98.3\%$ of full-CSI accuracy. These results suggest that substantial CSI feedback savings could be obtained using standardized reporting mechanisms without requiring learned CSI compression or reconstruction.

\vspace{-1. em}
%\IEEEtriggeratref{12}
\bibliographystyle{IEEEtran}
\bibliography{Bib_References/conf_short,
Bib_References/IEEEabrv,
Bib_References/Bibliography, Bib_References/multimodalsensing_Bio, Bib_References/ML_Bio, Bib_References/wifisensing}

\end{document}